# MISSION CONCEPT FOR THE SINGLE APERTURE FAR-INFRARED (SAFIR) OBSERVATORY

DOMINIC J. BENFORD[1], MICHAEL J. AMATO[2], JOHN C. MATHER[1]
S. HARVEY MOSELEY JR[1] and DAVID T. LEISAWITZ

[1]*Infrared Astrophysics Branch, NASA-Goddard Space Flight Center, Greenbelt, MD, U.S.A.;*
*E-mail: dominic.benford@nasa.gov*
[2]*Instrument Systems Branch, NASA-Goddard Space Flight Center, Greenbelt, MD, U.S.A.*

**Abstract.** The Single Aperture Far-InfraRed (SAFIR) Observatory's science goals are driven by the fact that the earliest stages of almost all phenomena in the universe are shrouded in absorption by and emission from cool dust and gas that emits strongly in the far-infrared (40 $\mu$m–200 $\mu$m) and submillimeter (200 $\mu$m–1 mm). In the very early universe, the warm gas of newly collapsing, unenriched galaxies will be revealed by molecular hydrogen emission lines at these long wavelengths. High redshift quasars are found to have substantial reservoirs of cool gas and dust, indicative of substantial metal enrichment early in the history of the universe. As a result, even early stages of galaxy formation will show powerful far-infrared emission. The combination of strong dust emission and large redshift ($1 < z < 7$) of these galaxies means that they can only be studied in the far-infrared and submillimeter. For nearby galaxies, many of the most active galaxies in the universe appear to be those whose gaseous disks are interacting in violent collisions. The details of these galaxies, including the effect of the central black holes that probably exist in most of them, are obscured to shorter wavelength optical and ultraviolet observatories by the large amounts of dust in their interstellar media. Within our own galaxy, the earliest stages of star formation, when gas and dust clouds are collapsing and the beginnings of a central star are taking shape, can only be observed in the far-infrared and submillimeter. The cold dust that ultimately forms the planetary systems, as well as the cool "debris" dust clouds that indicate the likelihood of planetary sized bodies around more developed stars, can only be observed at wavelengths longward of 20 $\mu$m.

Over the past several years, there has been an increasing recognition of the critical importance of the far-infrared to submillimeter spectral region to addressing fundamental astrophysical problems, ranging from cosmological questions to understanding how our own Solar System came into being. The development of large, far-infrared telescopes in space has become more feasible with the combination of developments for the James Webb Space Telescope (JWST) of enabling breakthroughs in detector technology.

We have developed a preliminary but comprehensive mission concept for SAFIR, as a 10 m-class far-infrared and submillimeter observatory that would begin development later in this decade to meet the needs outlined above. Its operating temperature ($\leq 4$ K) and instrument complement would be optimized to reach the natural sky confusion limit in the far-infrared with diffraction-limited performance down to at least the atmospheric cutoff, $\lambda \gtrsim 40$ $\mu$m. This would provide a point source sensitivity improvement of several orders of magnitude over that of the Spitzer Space Telescope (previously SIRTF) or the Herschel Space Observatory. Additionally, it would have an angular resolution 12 times finer than that of Spitzer and three times finer than Herschel. This sensitivity and angular resolution are necessary to perform imaging and spectroscopic studies of individual galaxies in the early universe. We have considered many aspects of the SAFIR mission, including the telescope technology (optical design, materials, and packaging), detector needs and technologies, cooling method and required technology developments, attitude and pointing, power systems, launch vehicle, and mission operations. The most challenging requirements for this mission are operating temperature and aperture size of the telescope, and the development of detector arrays. SAFIR can take advantage of much of the technology under development for JWST, but with much less stringent requirements on optical accuracy.

**Keywords:** far-infrared, submillimeter, telescope, SAFIR

## 1. Introduction

The Single Aperture Far-InfraRed (SAFIR) Observatory has been recommended by consensus of the broad astronomical community in the National Academy of Sciences Astronomy Decadal Report (NAS Decade Report, 2001) as a high priority scientific and technical successor to JWST and Spitzer (previously SIRTF). Additionally, the far-infrared community issued a more specific recommendation (Far-IR/SubMM Community Plan, 2003)

> The first step [of a unified plan to implement the Decadal Report's vision] is to develop the technology and start the planning for a cooled JWST class far-IR observatory called SAFIR (Single Aperture Far-IR telescope), to be operated like HST for a wide user community with a launch by the middle of the JWST lifetime in 2015.

This recommendation recognizes the exciting science opportunities offered by the promise of a dramatic increase in sensitivity and angular resolution of a facility like SAFIR in the far-infrared spectral region. SAFIR "will enable the study of galaxy formation and the earliest stage of star formation by revealing regions too enshrouded by dust to be studied by JWST, and too warm to be studied effectively with ALMA" (NAS Decade Report, 2001). In this paper, we present a mission concept for SAFIR as envisioned in a 10 m-class far-infrared observatory that would begin full-scale development late in this decade, for launch in ∼2015. For simplicity, we shall use the term *far-infrared* to denote the entire SAFIR wavelength range (20 $\mu$m–1 mm, Section 3) unless otherwise noted.

Advances in sensitivity and angular resolution in the far-infrared have lagged relative to the optical, near-infrared, and radio portions of the spectrum for two main reasons: the need to be above most or all of the Earth's atmosphere because of very strong absorption features of water vapor, and the need for very low temperature telescopes and instruments to reach the natural background limits of sensitivity at these wavelengths. In addition, large apertures are required to achieve the angular resolution needed for comparison with modest optical and near-infrared studies, and coherent interferometry (as is used in the radio to achieve high angular resolution without large single apertures) becomes increasingly challenging at high frequencies. As a result of efforts on other large space telescopes, strategies are now in hand to address effectively the simultaneous needs of a low temperature and a large single aperture.

The current decade will see the launch of three new observatory-class far-infrared facilities: Spitzer, SOFIA (the Stratospheric Observatory for Infrared Astronomy), and Herschel, each of which will provide substantial improvements in sensitivity and/or angular resolution as compared to facilities of the past such as the Kuiper Airborne Observatory (KAO), the InfraRed Astronomy Satellite (IRAS), the Cosmic Background Explorer (COBE), and the Infrared Space Observatory (ISO). Each of the present missions, however, is also clearly far from the current state-of-the-art in aperture and sensitivity for different reasons. Spitzer, a liquid helium cooled telescope launched in August 2003, will easily reach the confusion limit on the sky in the far-infrared, but because of its modest 85 cm aperture, this confusion limit is substantially above that of the other two facilities. SOFIA, which will begin science operations in 2005, will have three times the aperture diameter of Spitzer, but because it will operate at ambient stratospheric temperatures, it will never approach the sensitivity of cryogenic space observatories. Late in the decade the Herschel Space Observatory will provide improvements in both aperture (3.5 m) and operating temperature ($T \approx 80$ K). At the long end of the far-infrared spectrum, $\lambda \gtrsim 100\,\mu$m, Herschel will reach a much lower confusion limit on the sky than Spitzer or SOFIA, but with integration times of order a few hours and focal plane arrays of a few $\times 10^2$–$10^3$ pixels. Herschel too is handicapped because its operating temperature is still so much greater than the equivalent brightness temperature of the natural sky background at these wavelengths. While Spitzer will probe the faintest and most distant sources in the universe, SOFIA and Herschel will view the brighter sources with higher spatial and spectral resolution. Even before these new infrared observatories produce their stunning discoveries, it is clear that their scientific and technical trajectory points directly at the need of a capable, far-infrared optimized observatory like SAFIR.

SAFIR will use state-of-the-art technology in lightweight optics, cryogenic cooling, deployable structures, and the most recent detector technology to enable enormous sensitivity improvements over these previous missions, together with much needed improvement in angular resolution. For example, the mapping speed—which is proportional to pixels/sensitivity$^2$—will be $10^6$ times faster with two orders of magnitude more detectors and a two order of magnitude improvement in sensitivity. We touch on the compelling science drivers for this mission in the following section. Following that, we then discuss the drivers on telescope and detector specifications that result from the science requirements and our confidence in their achievability. We outline the concepts for the telescope design and discuss the ultimate sensitivity achievable by our approach. Following that, we present the mission design in detail. Finally, we briefly describe the current status of the project.

## 2. Science Drivers

We shall not provide a complete scientific justification for the SAFIR mission here, as it has been treated in other publications (Harvey et al., 2003; Rieke et al., 2003).

However, a few words of introduction are suitable, as the explicit mission requirements (Amato et al., 2003) were derived from the scientific questions outlined below.

2.1. OVERVIEW—THE ROLE OF THE FAR IR/SUBMM

Regardless of the original emission process, cosmic energy sources glow in the far-infrared and submillimeter. The broadband emission is due to the incredible efficiency of interstellar dust in absorbing visible and ultraviolet photons and reemitting their energy as thermal continuum emission. The appearance of the early Universe—of active galactic nuclei (AGN) and starbursting galaxies—and of star forming regions is transformed through this suppression of the visible and ultraviolet and corresponding augmentation of the far-infrared and submillimeter. At the same time, line emission in the far-infrared and submillimeter traces a variety of states of gas, from the heated cores of distant quasars to nearby cold protostellar clouds to our own Solar System's planetary atmospheres. Low-lying far-infrared fine structure lines are the major coolants for interstellar gas. Molecular transitions in this spectral range carry the signature of conditions in warm and dense interstellar clouds where stars and their solar systems form. Thus, we must look in the far-infrared and submillimeter for clues to the underlying processes shaping the origin, structure, and evolution of our Universe. We need large cold telescopes to reveal faint distant sources near the edge of the observable universe, and to show us the details of warm sources, even nearby, with clarity that matches our capabilities for seeing hotter material.

2.2. DISCOVERY OF NEW PHENOMENA

Technological advances enable astronomical discoveries. Harwit (1981) tried to quantify this relation in "Cosmic Discovery." In the 25 years preceding publication of the book, important discoveries were made in the first 5 years of the development of new technology that made them possible. The exceptional discovery potential in the far-infrared to submillimeter region arises because the sensors are still substantially short of fundamental performance limits for the natural sky background from space, and the telescopes available to date have been very modest in aperture (less than 1 m). The previous Decadal Report (*The Decade of Discovery in Astronomy and Astrophysics,* 1991) made use of a parameter, which they called "astronomical capability," to describe the discovery potential of new missions. This parameter is proportional to the time required to obtain a given number of image elements to a given sensitivity limit. SAFIR will have astronomical capability exceeding that of past far-infrared facilities by a factor of about $10^{10}$, and will still offer a gain of about $10^5$ after Spitzer and Herschel have flown. A gain of this magnitude is similar to the gain from the initial use of the Hooker 100-Inch Telescope on Mt. Wilson to the Hubble Space Telescope. The capabilities of SAFIR will enable enormous progress on some of the most important current problems in astronomy, and this huge gain in capability is certain to lead in new and exciting directions as well.

TABLE I
Telescope requirements for SAFIR

| Parameter | Requirement | Science targets |
|---|---|---|
| Aperture | >8 m | High $z$ galaxies; Debris Disks |
| Temperature | $\sim 4$ K | L* Galaxy at $z \sim 5$; zodi limit |
| Wavelength range | 20–800 $\mu$m | JWST overlap; Gas cooling lines |
| Diffraction limit | $\geq 40\,\mu$m (1″) | Debris disks, distant galaxies |
| Pointing accuracy | $\sim 0.5 - 1.0''$ | Driven by 40 $\mu$m Diffraction limit |
| Pointing stability | $\sim 0.1''$ | Driven by 40 $\mu$m Diffraction limit |
| Lifetime | >5 Years | Productivity |

## 3. Telescope and Instrument Requirements

Table I summarizes the top level requirements on the SAFIR telescope and the science observations driving each of them.

The two most critical requirements are aperture and operating temperature, assuming the emissivity is minimized to a level limited by optical coatings and reasonable stray light considerations. The aperture and telescope background emission determine its ultimate limiting sensitivity (see Section 5) as well as the practical limit of the number of seconds of integration required to reach the confusion limit on the sky. The science drivers for SAFIR require the confusion noise to be low enough to observe typical dusty L* galaxy formation close to the epoch of metal enrichment, $z > 5$. Another important driver is the ability to study the formation and evolution of planetary systems via their debris disks out to modest distances, D $\sim 100$ pc. These requirements imply an aperture of order 10 m for SAFIR. Figure 1 shows an estimate of the confusion limit for broadband observations with a 10 m aperture telescope (Blain et al., 1998) as a function of wavelength relative to the photon background limited sensitivity for an assumed 5% emissive telescope of various possible temperatures and for two representative integration times, 1 and $10^4$ s. To reach a sensitivity and confusion limit somewhat less than the brightness of L* galaxies at $z \sim 5$ at 100 $\mu$m ($\sim 4\,\mu$Jy) requires $T \leq 10$ K. For spectroscopic purposes, however, the confusion limit is much lower and it is therefore important that the telescope be even colder, 4 K, to detect lines from distant galaxies.

The next most critical requirement is on the short wavelength limit for SAFIR, because this drives the optical surface and alignment tolerances and thus the technology chosen for the primary mirror. In order to provide high sensitivity and angular resolution for complete coverage of the very wide range of wavelengths in the infrared, SAFIR's wavelength coverage would extend down to overlap with JWST's at $\lambda \sim 20\,\mu$m. The science drivers described above require diffraction-limited performance at a somewhat more relaxed wavelength limit of 40 $\mu$m for imaging of debris disks and distant galaxies. The diffraction limit at 40 $\mu$m for a 10 m telescope is $\approx 1''$ (1.22 $\lambda/D$). Therefore, the requirements on pointing accuracy and stability ($\sim 0.5 - 1''$ and $0.1''$, respectively), as well as surface quality for SAFIR will be relatively modest in comparison with JWST. Thus the JWST engineering research and development can be assumed to carry over in part to SAFIR. In the post-JWST era, the biggest technological challenge is likely to be the lower required telescope temperature.

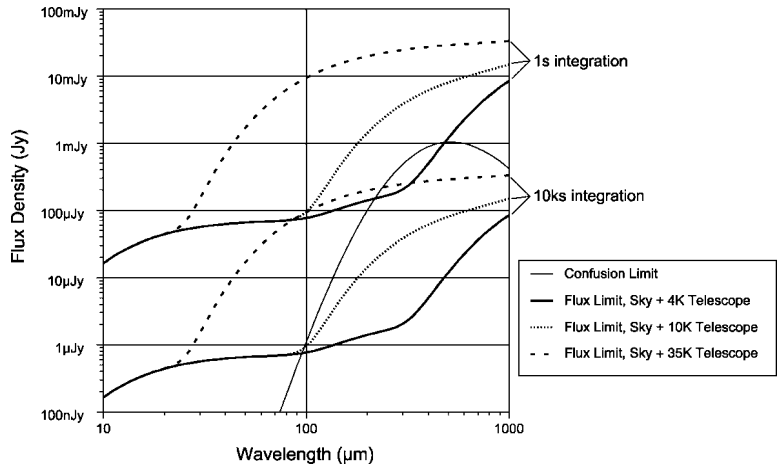

*Figure 1.* SAFIR sensitivity as a function of telescope temperature and integration time, as compared to the confusion limit.

The science program that motivates the need for SAFIR requires a complement of imaging and spectroscopic instruments over the entire wavelength range of SAFIR from 20–800 $\mu$m. Table II lists a strawman complement of instruments, noting the spectral resolution ($R \equiv \lambda/\Delta\lambda$) and the field of view, along with some of the science drivers from Section 2.

The strawman camera uses a dichroic and a double filter wheel to provide simultaneous diffraction-limited imaging in two bands, one in the 20–100 $\mu$m range with 1' field-of-view, the other from 140–600 $\mu$m with 4' field-of-view. For the low resolution spectrometer, a small field image slicing spectrometer is well matched to the size of many of the objects being studied, and provides instantaneous full-wavelength coverage by using a pair of broadband gratings. For the moderate resolution $R \sim 2000$ spectrometer, a long slit grating spectrometer is necessary, although at the shorter wavelengths this could be achieved with an image slicing spectrometer similar in design to FIFI-LS on SOFIA (Looney et al., 2003). Given current and near-future technology developments, the high resolution spectrometer will likely be a single-moded coherent receiver. For this instrument, the development of extremely broad frequency range detectors (such as hot electron bolometers) and local oscillators.

TABLE II

Strawman instrument complement for SAFIR

| Instrument | Wavelength range<br>Driving science | Spectral resolution | FOV |
|---|---|---|---|
| Camera | 20–600 $\mu$m | ∼5 | 1–4$'$ |
| | High $z$ Reddenning; searches for KBOs; imaging surveys... | | |
| Spectrometer | 20–100 $\mu$m | ∼100 | ∼10$''$ (Image slicing) |
| (low resolution) | Ice features in Debris Disks, YSOs, etc. | | |
| Spectrometer | 20–800 $\mu$m | ∼2000 (150 km/s) | ∼1$'$ |
| (moderate resolution) | Cooling lines (e.g., $C^+$, $N^+$); Chemical evolution | | |
| Spectrometer | 25–520 $\mu$m | ∼$10^6$ (0.3 km/s) | >1 beam |
| (high resolution) | Dynamics of star forming regions, evolved stars; Gas Cooling | | |

Figure 2 displays the sensitivity of SAFIR relative to a number of precursor missions as well as several that are hoped to follow it. Also shown in Figure 3 is the angular resolution of SAFIR as a function of wavelength in the context of existing, planned, and possible future missions. It is clear from these figures that SAFIR not only offers a dramatic increase in sensitivity over precursor missions, but provides a bridge in sensitivity between what will be the most powerful telescope at shorter wavelengths (JWST), and that at longer wavelengths (ALMA).

## 4. Telescope Concepts and Expected Performance

SAFIR has been recommended by several assemblies (NAS Decade Report, 2001; Far-IR/SubMM Community Plan, 2003) during the past few years. They have made these recommendations because information vital to the realization of NASA's major scientific objectives is uniquely available in the far-infrared. It is important to recognize that SAFIR is the first and foremost mission to fulfill this scientific need. Further study of SAFIR mission design concepts will be needed before a single approach that accomplishes the highest priority science goals with ready technology, subject to programmatic considerations, can be selected. Three possible concepts are shown below in Figure 4: (top) based on the original JWST, the "NGST Strawman" design; (middle) derived from the current JWST for maximum heritage and fidelity; (bottom) based on a sparse aperture telescope to improve angular resolution. The sparse aperture concept has the equivalent collecting area of the two more compact telescopes, and is 17 m tip-to-tip.

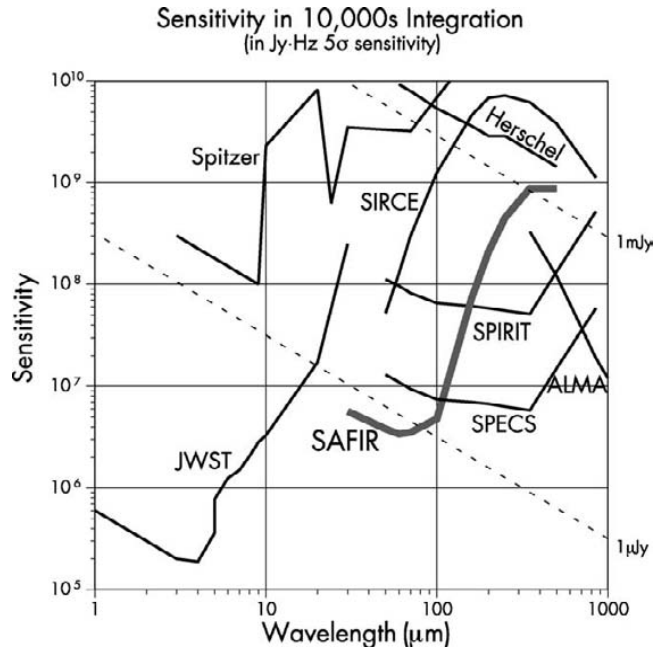

*Figure 2.* SAFIR sensitivity compared to present and future mission concepts.

The SAFIR mission shown at the top in Figure 4, and detailed in Figure 5, is a mission concept that has been studied in detail at NASA's Goddard Space Flight Center. It is based on the National Academy's recommendation: "To take the next step in exploring [the far-infrared], the committee recommends the SAFIR Observatory, a... telescope that builds on the technology developed for JWST." (NAS Decade Report, 2001).

Taking the contemporary (2002) JWST design as a starting point, a detailed analysis was made of the changes necessary to produce a SAFIR that is implemented as a larger, colder version of JWST (Amato et al., 2003). A broad list of science investigations was used to generate an explicit list of technology requirements, from which the observatory requirements (Section 3) are derived. After the selection of the JWST prime contractor, a second design more directly emulating the chosen JWST architecture was developed (Figure 4, middle); a further design with a sparse aperture telescope was adapted from this architecture (Figure 4, bottom).

One of the immediate differences between SAFIR and JWST is that in order to achieve the ultimate sensitivity for the difficult spectroscopic observations planned for SAFIR, the telescope and other optics will have to be cooled to 4 K, well below the ∼35 K achieved by JWST's passively cooled architecture. GSFC's conceptual design for SAFIR uses cascaded cryocoolers to provide moderate cooling powers at 40 K, 15 K, and 4 K. The JWST-like sunshade is mounted on the 40 K stage, whereas a single additional layer of the sunshade is mounted on the 15 K stage.

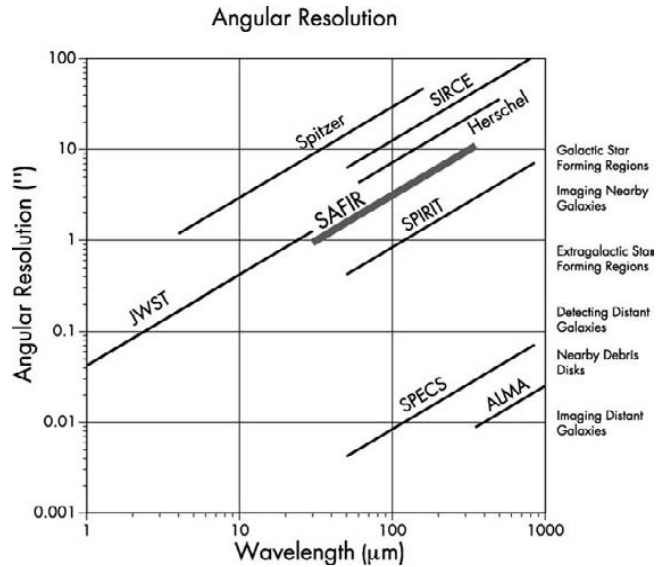

*Figure 3.* SAFIR angular resolution compared to present and future mission concepts.

This sunshade provides an environment so well shielded from the Sun that only modest cooling is needed to cool the telescope to 4 K.

For JWST, there were two methods developed for deploying a large, rigid primary mirror: petal-like folding or drop-leaf-table-like folding. The GSFC SAFIR design baseline is to reproduce JWST's approach, using a table folding. Other SAFIR options—particularly for a sparse aperture (Figure 4, bottom)—would use a petal fold. Since SAFIR is larger than JWST, the designs will diverge somewhat: a petal design will result in small notches around the outside of the aperture, while a table-fold design may have two small slices off the aperture edges. The PSF will be slightly degraded, but the collecting area at large radii will still benefit confusion-limited and sensitivity-limited observations. The choice of optical surface is largely independent of the method of deployment and can be deferred until JWST has validated its technology. JWST has selected a C/SiC mirror surface with a composite carbon fiber structure. Among the other materials under consideration for SAFIR are carbon fiber mirrors and structure (possibly with glass face-sheets), C/SiC mirrors and structure, beryllium, and aluminum mirrors. Given that SAFIR does not require diffraction-limited performance at $\lambda < 40\,\mu$m, it might be possible to duplicate the JWST mirror technology, but without the final polishing process. It is also possible that the telescope could use a non-JWST-like optical design, as discussed below.

To fit the four instruments outlined in Table II, a large field of view ($\sim 15'$ diameter) is needed. Because the diffraction-limited resolution at the shortest wavelengths is $\simeq 1''$, a pointing stability of $0.1''$ is desirable. This can be achieved only with a cryogenic pointing camera located on the instrument package. Our concept uses a

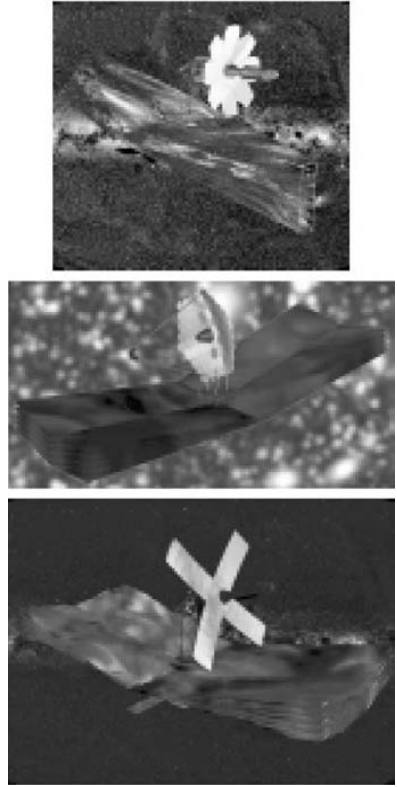

*Figure 4.* A set of possible SAFIR mission concepts developed at NASA/GSFC. From top to bottom, these are the "NGST Strawman" design, the "JWST Heritage" design, and the "Sparse Aperture" design.

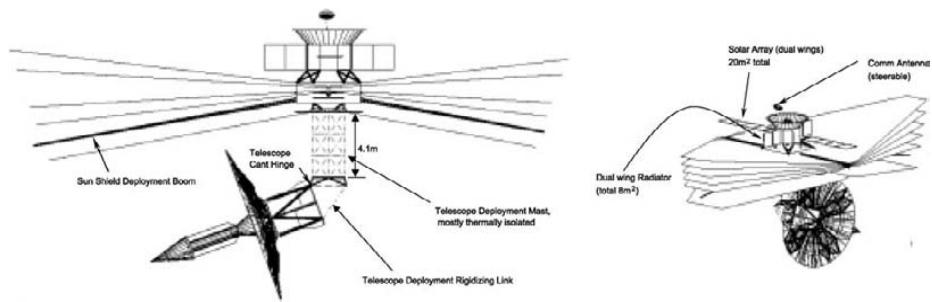

*Figure 5.* SAFIR concept developed at NASA/GSFC, based on JWST heritage.

2 $\mu$m focal plane camera with a large field of view and high speed subarray readout to measure jitter offsets from known 2MASS point sources.

There is work underway to develop a new architecture to break the paradigm of launching large, often massive mirrors into space. NASA has been developing a concept using membrane mirrors tensioned by truss structures assembled by astronauts. Called the Dual Anamorphic Reflector Telescope (DART; Dragovan, 2002), this telescope consists of two convex parabolic cylindrical reflectors oriented with respect to each other to produce a point focus (Figure 6). Since each reflector contains only a single simple curve, the mirrors can be formed by tensioning a reflective foil over a frame with a parabolic contour along one axis. The use of an extremely low-mass membrane for the reflective surfaces would significantly reduce the mass of the telescope support. The DART architecture uses a thin membrane for its reflectors, so the density of the mirror surface is independent of size. In fact, the ratio of structure to reflector mass can decrease with increasing aperture size. The approach has significant implementation issues for use in SAFIR; for instance, it may be most effectively assembled by astronauts in orbit, and the large, sparse membranes and structure may be more difficult to cool to 4 K than a more compact conventional mirror technology. If the problems of deploying and cooling a membrane mirror can be solved, and the field of view and surface figure made adequate, this technology might be a viable replacement for the JWST mirror in a SAFIR concept.

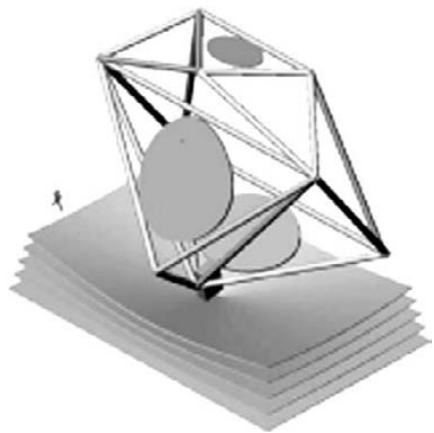

*Figure 6.* DART concept.

## 5. Ultimate Sensitivity of a Cryogenic Far-Infrared Telescope

It is reasonable to assume that a future space-based far-infrared mission will likely require that the system sensitivity be maximized for point source detection. The ideally sensitive instrument will be limited only by the photon statistics from the natural backgrounds outside of Earth orbit. The natural sky background for a cold telescope in space is extremely low as compared to the background from a warm telescope or beneath the Earths atmosphere. To take advantage in this limit, instrument sensitivity can be increased by orders of magnitude.

The natural sky background limit is set by the combination of several components (Figure 7), and is particularly low in the 50–500 $\mu$m regime. In order to keep below this intensity limit, the telescope must be cold: 4 K to be equal to the sky background (Figure 8). An optimally sensitive telescope for operation in the far-infrared such as SAFIR must be this cold.

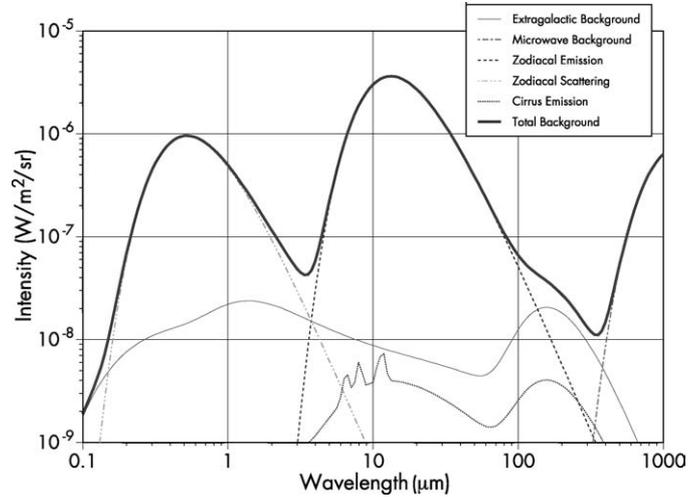

*Figure 7.* Intensity of natural sky background in a diffraction limited beam of unity bandwidth; the region from 50–500 $\mu$m is particularly dark.

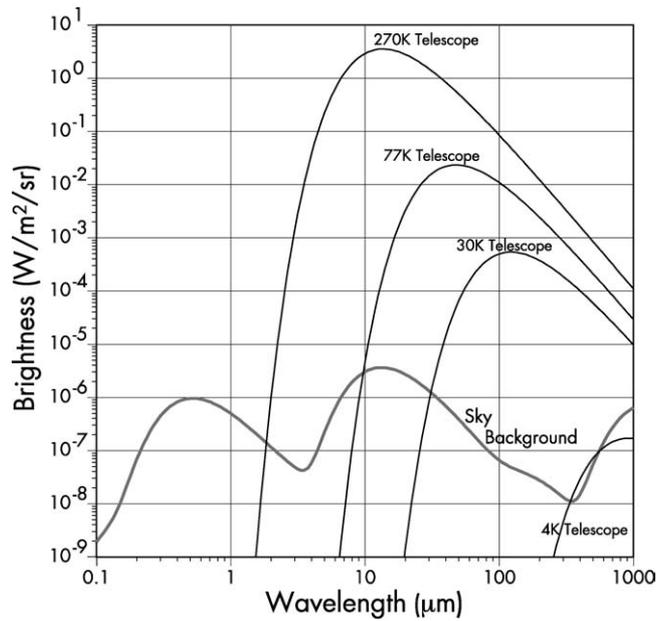

*Figure 8.* The total sky background compared with the emission of telescopes of 5% emissivity.

Of course, there are other limits to ultimate sensitivity besides the photon noise limit presented above. In the case of broadband observations in the far-infrared, the confusion limit is significant. The MIPS instrument on Spitzer is expected to be confusion limited (one source detected at the $1\sigma$ limit per beam) at 160 $\mu$m in only seconds of integration. The confusion limit drops as a very fast function of the mirror diameter, due to the increased resolving power. The tradeoff wavelength between photon-limited (where collecting area is important) and confusion-limited (where diameter is important) sensitivity is a function of the diameter, and can be extrapolated to future space telescopes. Using the photon noise calculated above with the confusion limit calculated by Blain et al. (1998) we find that this wavelength falls between 30 $\mu$m for a 1-m-class telescope and 80 $\mu$m for a SAFIR-sized observatory (Figure 9).

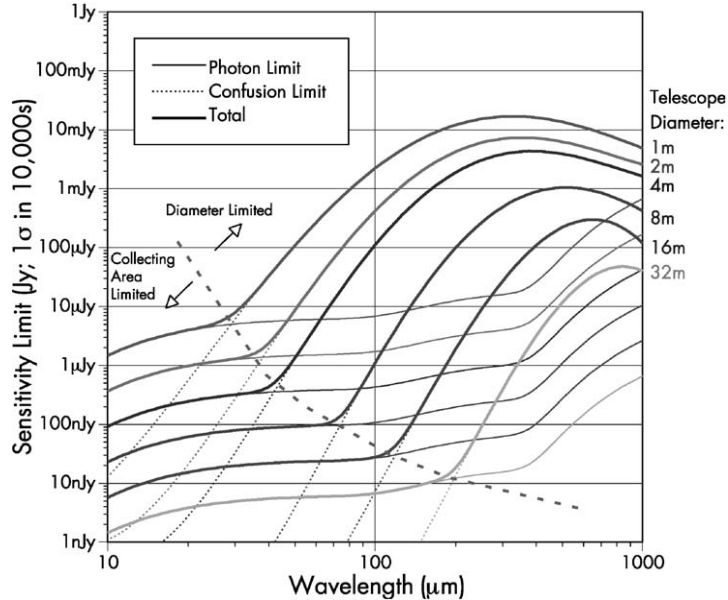

*Figure 9.* Sensitivity limits of a cryogenic space observatory, showing the balance between photon noise (collecting area) limited and confusion (diameter) limited observations.

In the case of narrower bandwidth observations, the situation changes: the confusion limit to sensitivity is noticeably reduced. As a simple approximation, we calculate the sensitivity of SAFIR with a spectral resolution of $R = \lambda/\Delta\lambda = 1000$, and assuming no confusion. In Figure 10 (Left), we compare this sensitivity with different telescope temperatures. We then calculate the worst case ratio of sensitivity degradation at a given temperature. This resultant sensitivity loss can be considered as an effective aperture loss. This is plotted in Figure 10 (Right), for cases with confusion ($R = 5$) and without confusion ($R = 1000$). In reality, the confusion limit at $R = 1000$ is not zero, but around a factor of ∼100 below the limit at $R = 5$. This level of confusion is sufficiently low and so it does not significantly limit the sensitivity until integration times of order $10^5$ s are considered.

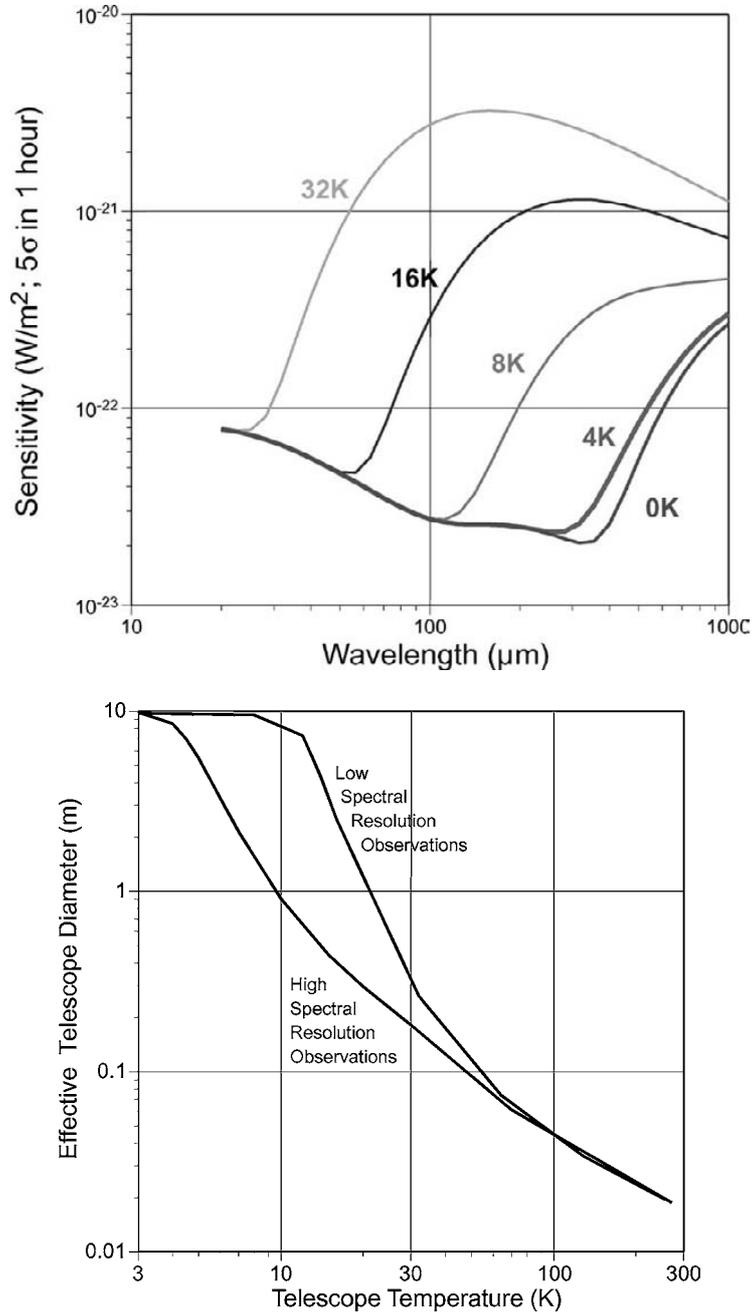

*Figure 10.* (Top) Sensitivity of SAFIR at a variety of telescope temperatures, assuming 1% emissivity and a spectral resolution of $\lambda/D\lambda = 1000$ (300 km/s). (Bottom) Degradation in effective aperture as telescope temperature increases, for 1% emissivity and the cases of a spectral resolution of $\lambda/D\lambda = 5$ with confusion and for $\lambda/D\lambda = 1000$ with low confusion.

## 6. SAFIR Mission Design

The National Academy has recommended that a SAFIR design draw on JWST technology (NAS Decade Report, 2001). Because this is the first published SAFIR mission design, building on JWST provides the highest fidelity design on the shortest timescale, and is a suitable point of departure for other designs. A 10-m diameter, 4 K telescope with a far-infrared instrument suite of three spectrometers and a broadband camera is challenging. To mitigate this challenge, we consider what of the JWST design and technologies might be reused for the SAFIR observatory. This is a necessary exercise to pursue to show the feasibility of SAFIR and to lend credibility to the design, with its estimated parameters, cost, and schedule. We have developed a SAFIR conceptual design that is heavily based on many JWST designs. We think this approach makes sense because JWST will spend a great deal of time and money making its design and technologies work several years in advance of SAFIR. It is expected that this approach has the potential to reduce SAFIR development time and costs significantly.

Because our current baseline SAFIR concept is an JWST-like concept, we must consider needed technologies and technology development with an eye on what JWST will already need to have completed prior to its launch. In this section, we present a description of the SAFIR conceptual design (shown in Figure 11), highlighting needed technologies, their development status today and the needed status prior to JWST launch. Several technologies—most notably the detectors and cryocoolers—will require developments specific to SAFIR and future far-infrared missions.

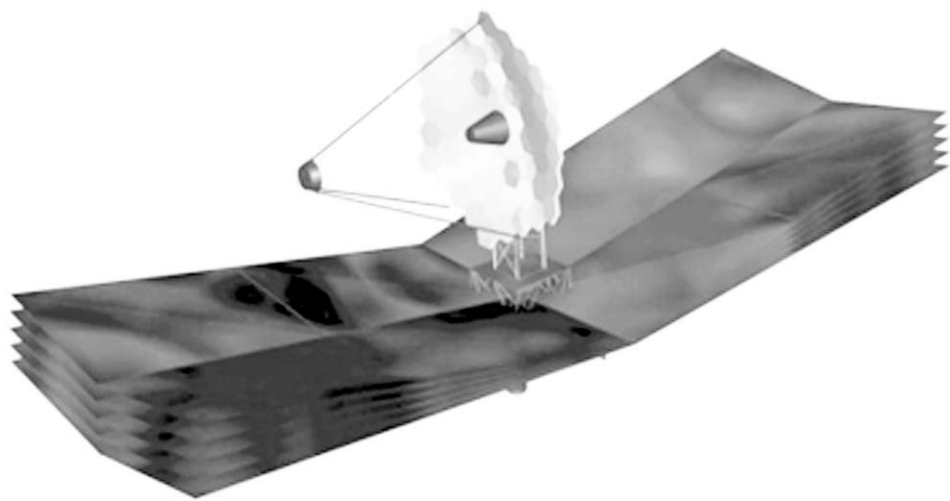

*Figure 11*. SAFIR concept based on JWST; developed by NASA/GSFC.

6.1. TELESCOPE DESIGN

The telescope is the most fundamental part of a large observatory. If the premise of an initial SAFIR concept is to take advantage of any designs and technologies of JWST that make sense for SAFIR, then the SAFIR architecture should use a hinged folding primary mirror. However, SAFIR requires twice the collecting area of JWST, and so a petal-folding primary mirror may be the only suitable approach, even if it does not draw on JWST heritage. However, the concepts are not substantially different for our purposes and both can be used as a basis for a SAFIR mission.

The three major SAFIR requirements that differ from JWST parameters are the size of the primary mirror, the colder temperature, and the longer wavelength range of operation. With the exception of the caveat mentioned above, both proposed JWST approaches can be used to deploy a larger telescope. The basic SAFIR concept uses designs not greatly different from current JWST concepts to deploy the larger SAFIR telescope.

6.2. TELESCOPE TECHNOLOGIES

The telescope requires development in three distinct technology areas to realize a lightweight, cryogenic 10 m observatory. For its purposes, most of these will be addressed by JWST developments. First, the method of folding and deploying the primary mirror will require careful mechanical design and substantial ground testing. SAFIR should use the same approach as JWST, but with a larger primary mirror diameter. If the mirror is segmented into leaves—the baseline approach—it will be constrained to a maximum size of around 9 m, but with a completely filled aperture. If the mirror is segmented into petals, a 10-m diameter mirror can be stowed into a launch vehicle fairing, but with notched regions around the outer edge.

The second technology involved is the method of phasing and controlling the mirror surface, which requires sensitive detection of position errors and precision actuation of the segments. SAFIR's longer wavelengths make this task roughly 10 times less precise than that for JWST, and the lower temperature will reduce thermal distortions that require JWST's mirror to be rephased periodically. It is likely that the mirror will need to be phased only once and left for the mission duration, but the actuators will probably still be needed as a contingency measure.

The third technology is the mirror surface material. Several options have been investigated for JWST, including glass face-sheeted composites, C/SiC, and Beryllium. SAFIR may be able to use polished aluminum as well, if the figure can be made smooth enough. JWST has chosen C/SiC, a technology that works well at $\sim$35 K and is lightweight but stiff. SAFIR needs adequate thermal conductivity at 4 K, and would prefer lower areal density materials at the expense of some stiffness. The C/SiC approach will need to be investigated further for the SAFIR-specific implementation.

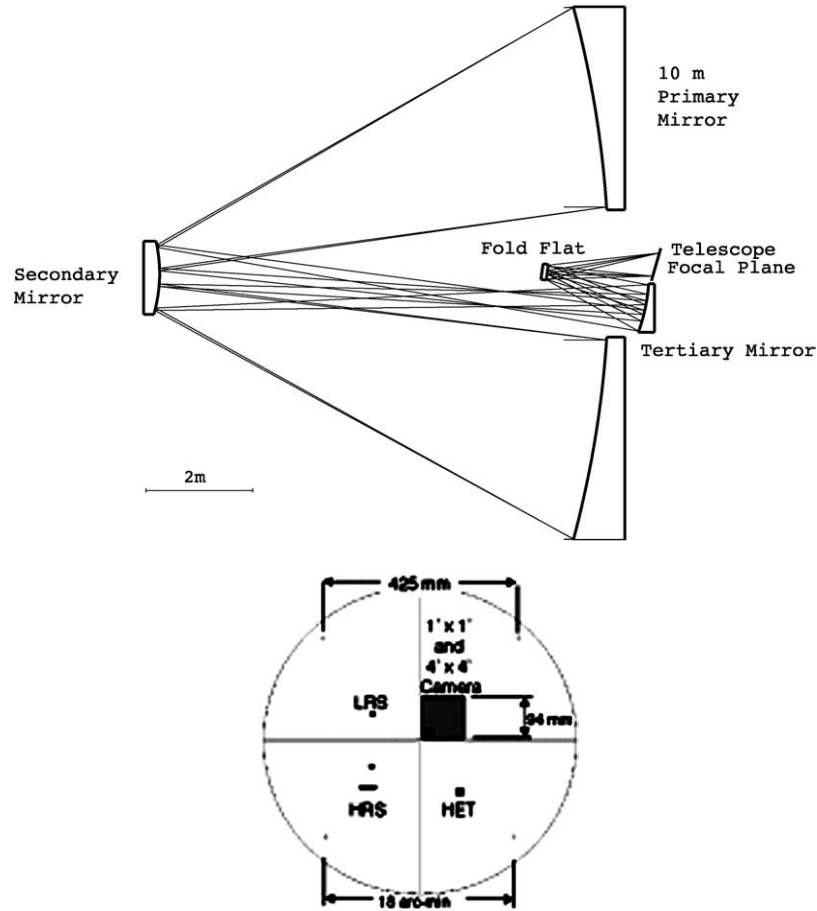

*Figure 12.* (Top) Optical layout of SAFIR telescope. (Bottom) focal plane arrangement of instruments.

Several optical designs for the telescope are possible, but we are using an JWST-like three mirror telescope as a baseline, shown at left in Figure 12. This design is slightly off-axis, but features a flat steering mirror that can be used for fine pointing correction (which may be needed, as mentioned in Section 6.8). The field of view of this design is large, permitting ample space for the four instruments (Figure 12, Right) and a focal plane star sensor (not shown).

A point spread function (PSF) map has been produced for the two different SAFIR designs. In Figure 13, the PSF of a circular SAFIR aperture is shown. In Figure 14, the PSF for the sparse aperture concept is shown. It is clear that the artifacts are more prominent for the sparse aperture version, but it also yields a narrower PSF (at $\lambda = 40\,\mu m$, $\theta \sim 0.6''$ as opposed to $\theta = 1.0''$) and has several low sidelobe regions; practically, the two PSFs are indistinguishable at $4''$ radius.

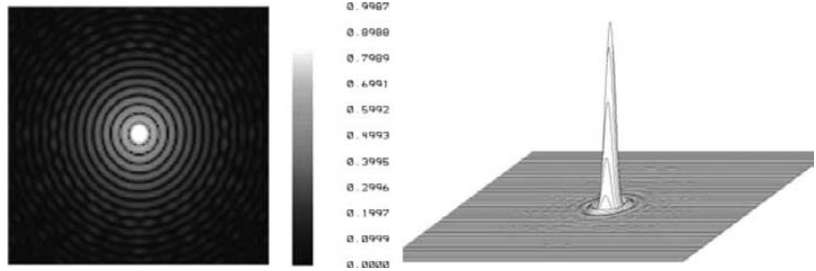

*Figure 13.* PSF of the circular aperture: (Left) log scale; (Right) linear scale.

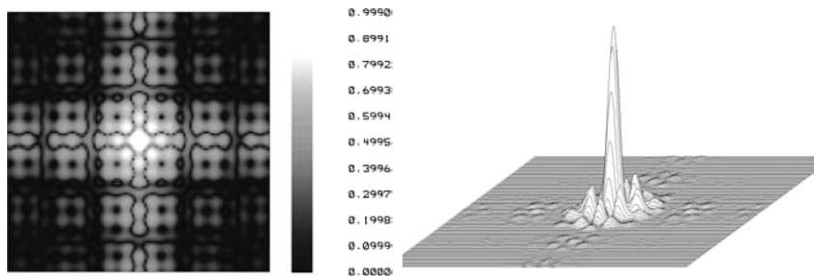

*Figure 14.* PSF of the sparse aperture: (Left) log scale; (Right) linear scale.

### 6.3. INSTRUMENT DESIGN AND DETECTOR NEEDS

To cover the wide range of wavelengths and spectral resolutions, SAFIR may need four instruments, summarized below:

- Instrument A; CAM: Broadband camera with spectral resolution of $R \sim 5$ covering 20–600 $\mu$m, using simultaneous two-wavelength imaging with a $1'$ field of view (20–100 $\mu$m) and a $4'$ field of view (140–600 $\mu$m). A filter wheel with 6 positions will permit the selection of any band over the entire range.
- Instrument B; LRS: Low resolution spectrometer with spectral resolution of $\sim$100 covering 25–100 $\mu$m, using two simultaneous integral field grating spectrographs. The field of view will be about $6''$ on a side from 25–50 $\mu$m and $12''$ on a side from 50–100 $\mu$m.
- Instrument C; MRS: Moderate resolution spectrometer with spectral resolution of $\sim$2000, covering 20–720 $\mu$m. At short wavelengths (20–120 $\mu$m), HRS features an integral field grating spectrometer with an $18''$ field of view, whereas at longer wavelengths (120–720 $\mu$m), it will use a long slit grating spectrometer with an adjustable slit up to $18'' \times 96''$ in size.
- Instrument D; HRS: High resolution spectrometer with spectral resolution of $\sim$100 000 covering 25–520 $\mu$m. A heterodyne spectrometer is the most likely technology for this application, and would have a sparsely-sampled array of single-moded detectors with at least $2 \times 2$ format, each pixel being one diffraction-limited beamwidth. A beam steering mirror is used to fill in the gaps of the array.

It is unlikely that any JWST detector-related technology can be used for SAFIR with the exception of a 2 $\mu$m camera for star tracking. To realize these instruments, the following technological investments need to be developed specifically to enable SAFIR:

- For CAM, large format arrays covering the 20–600 $\mu$m range are needed. Formats of $128^2$ pixels, each with a noise equivalent power (NEP) of $3 \times 10^{-19} \text{W}/\sqrt{\text{Hz}}$ are required. Superconducting bolometer arrays can, with sufficient development funding, meet these requirements.
- For MRS, large format arrays covering the 20–720 $\mu$m range are needed. Formats of $64^2$ pixels, each with a noise equivalent power (NEP) of $1 \times 10^{-20} \text{W}/\sqrt{\text{Hz}}$ are required. It may be necessary to develop novel approach to detectors to meet this sensitivity requirement.
- For LRS, similar large format arrays to MRS must be developed, but with lower sensitivity for higher photon background.
- For HRS, single-moded coherent detectors operating near the quantum limit over the 25–520 $\mu$m range (0.6–12 THz) are needed. To this end, increasing the tunable bandwidth of detectors and local oscillator sources is a key research investigation to be undertaken. Hot electron bolometers are the most promising current technology for this purpose, if they can be manufactured with near-quantum noise. Additionally, backend spectrometers using low-power digital autocorrelators should be developed.

### 6.4. DETECTOR TECHNOLOGIES

The far infrared and submillimeter ranges have benefited relatively little from investments in detector technology for nonastronomical pursuits. In this regard, they differ dramatically from the visible, near- and mid-infrared, and radio regions. Detectors in those spectral regions closely approach theoretical performance limits. For example, in the visible, CCDs have quantum efficiencies greater than 90%, read noises of about two electrons, and formats up to many millions of pixels. In the far-infrared the investment in technology to date has been modest, due to both technological hurdles in using the technology which have now largely been overcome, and the lack of direct applications in defense and commercial industry. This much smaller prior investment has left the possibility for orders of magnitude near-term progress toward fundamental limits. NASA missions are the only major customers for this technology, and an augmented NASA investment will be necessary to enable SAFIR and other far-infrared and submillimeter missions, including Explorers. These investments will guarantee our nation's leadership in this important technology.

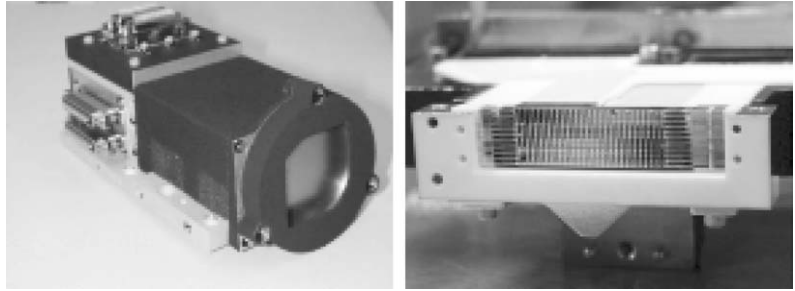

*Figure 15.* State-of-the-art large format far-infrared detector arrays: (Left) Spitzer/MIPS (Young et al., 2003); (Right) SHARC II (Dowell et al., 2003).

SAFIR requires sensitive arrays containing thousands of detectors. Far-infrared photoconductors are the most advanced in array construction, the best example of which is the $32 \times 32$ element $70\,\mu$m Spitzer/MIPS array (Figure 15 at left; Young et al. (2003). However, they fall somewhat short of theoretical limits in potential performance and respond only up to the excitation energy (around $\lambda_c = 200\,\mu$m). Development for SAFIR will require both larger arrays ($128 \times 128$) and longer wavelengths ($720\,\mu$m). Bolometers have broad spectral response and are the most advanced submillimeter continuum detectors, but they require extremely low operating temperatures. The largest existing bolometer array is the $12 \times 32$ array in the SHARC II instrument at the Caltech Submillimeter Observatory (Figure 15 at right; Dowell et al., (2003)—and its JFET readout technology cannot presently be multiplexed to permit larger arrays. Development needs to emphasize improved array technology, such as SQUID-based multiplexing, and superconducting-thermometer bolometers that interface well to SQUID electronics. Hot electron bolometer (HEB) mixers provide the best heterodyne operation above the superconducting gap frequency of NbTiN, at around 1200 GHz. Heterodyne detectors have large practical advantages for spectroscopy over photoconductors and bolometers. Development needs to address reducing noise temperatures and developing support electronics to allow large scale spatial arrays, and to push to higher frequencies and greater tunable frequency ranges.

### 6.4.1. *Detectors for Low Spectral Resolution*

The two lower spectral resolution instruments listed in Table III require very similar detectors. They differ slightly in the number of pixels, but more significantly in their noise equivalent power (NEP). We evaluate the detector requirements using the Richards figure of merit (Richards, 2003), the number of pixels divided by $NEP^2$ (which is essentially the mapping speed). We find that the requirement for the broadband camera is $1.6 \times 10^{42}$ pixels/W$^2$/s; for the low resolution spectrometer, it is $1 \times 10^{43}$ pixels/W$^2$/s. By way of comparison, the largest bolometer array in existence (SHARC II) has a figure of merit of about $4 \times 10^{36}$ pixels/W$^2$/s. (More sensitive detectors can be made, but the background for suborbital telescopes

TABLE III

SAFIR observatory conceptual design mass estimates

|  | Estimated Mass (Conservative) (kg) | Mass w/Contingency (kg) |
|---|---|---|
| PAYLOAD TOTAL | 2430 | 2916 |
| SPACECRAFT BUS TOTAL | 1403 | 1684 |
| Observatory Dry Mass | 3833 | 4600 |
| PROPELLANT | 260 | 311 |
| **TOTALS (BUS + PAYLOAD)** | **4093** | **4911** |
| Delta IV Heavy capability to L2 | 9410 | 9410 |
| Margin [kg] | 5317 | 4499 |
| Margin [%] | 56.5 | 47.8 |

is high enough that this is unnecessary.) Richards law states that, averaged over long times, the figure of merit of the state-of-the-art bolometers doubles every year, so that about 20 years of development ought to be needed to achieve the leap to the SAFIR broadband detectors. However, the revolution represented by superconducting bolometers is poised to permit an improvement in the state-of-the-art in the form of the SOFIA/SAFIRE detector array, which, when completed in 2005, will have a figure of merit of $5 \times 10^{40}$ pixels/W$^2$/s. From this starting point, Richards law suggests that the SAFIR detectors will be feasible for flight by 2013.

Using the background power calculation from Figure 7, we compute the power absorbed by a diffraction-limited detector on a 4 K, 10-m telescope. This power $P$ can be converted into a photon-limited noise equivalent power at optical frequency $\nu$ using $NEP_{\text{photon}} = \sqrt{2Ph\nu}$ and a photon-limited noise equivalent flux density (NEFD) assuming optical efficiency $\eta$ using $NEFD = 2NEP/(A\eta\Delta\nu)$. The results are shown for a spectral resolution of $\lambda/\Delta\lambda = 5$ and $\lambda/\Delta\lambda = 100$ in Figure 16. In the core SAFIR observing bands of 30–300 $\mu$m, the optical power is surprisingly flat, and the NEP values for the detector are $10^{-19}$ W/$\sqrt{\text{Hz}}$ at $\lambda/\Delta\lambda = 5$ and $2 \times 10^{-20}$ W/$\sqrt{\text{Hz}}$ at $\lambda/\Delta\lambda = 100$.

### 6.4.2. *Detectors for High Spectral Resolution*

Independent of optical approach, a maximum-sensitivity spectrometer will be dispersive. A large array of close-packed detectors is a useful thing in such a case, either to cover larger fields (e.g., for a Fabry-Perot) or more bandwidth (e.g., for a grating spectrometer). These detectors will have to be extremely sensitive, with NEPs of around $3 \times 10^{-21}$ W/$\sqrt{\text{Hz}}$. This is shown graphically in Figure 17, where the same calculations as discussed in Section 6.4.1 are applied for a resolving power of 2000 (equivalent to a velocity resolution of 150 km/s, suitable for minimally resolving lines in galaxies).

When building a high resolution spectrometer for a wavelength longward of $\sim$300 $\mu$m (1 THz), it is common to use coherent spectrometers. If the instantaneous bandwidth desired is higher than $\lambda/\Delta\lambda = 10^4$ (30 km/s), the bandwidth of the intermediate frequency amplifier and backend spectrometer need be only ~100 MHz, easily achievable by today's standards. Given sufficient technology investment, could a coherent spectrometer be developed to cover the SAFIR bands (20 $\mu$m–520 $\mu$m) to provide information via very high resolution observations? For the case of Galactic observations, where line strengths are relatively larger though line widths are narrow, this is possible: SAFIR's detection limit in a 1 km/s line in $10^4$ s is about 1 mK at 300 $\mu$m with a heterodyne spectrometer. However, in distant extragalactic sources where linewidths are larger but sources are fainter, it becomes harder for such an instrument due to the limitation of sensitivity from quantum noise. Figure 17 (right) shows a comparison between the NEFD of a quantum limited coherent spectrometer and a high resolution dispersive direct detection spectrometer, both with $\lambda/\Delta\lambda = 10^4$. To put the comparison into scientific terms, the predicted fluxes of bright far-infrared fine-structure lines from starburst galaxies as they are redshifted to submillimeter wavelengths are shown. The quantum limit will make the detection of these lines at $z > 1$ a difficult proposition. In the quantum limited case, a 270 K telescope is as good as a 4 K telescope, and so ALMA would be more sensitive than SAFIR wherever they overlapped.

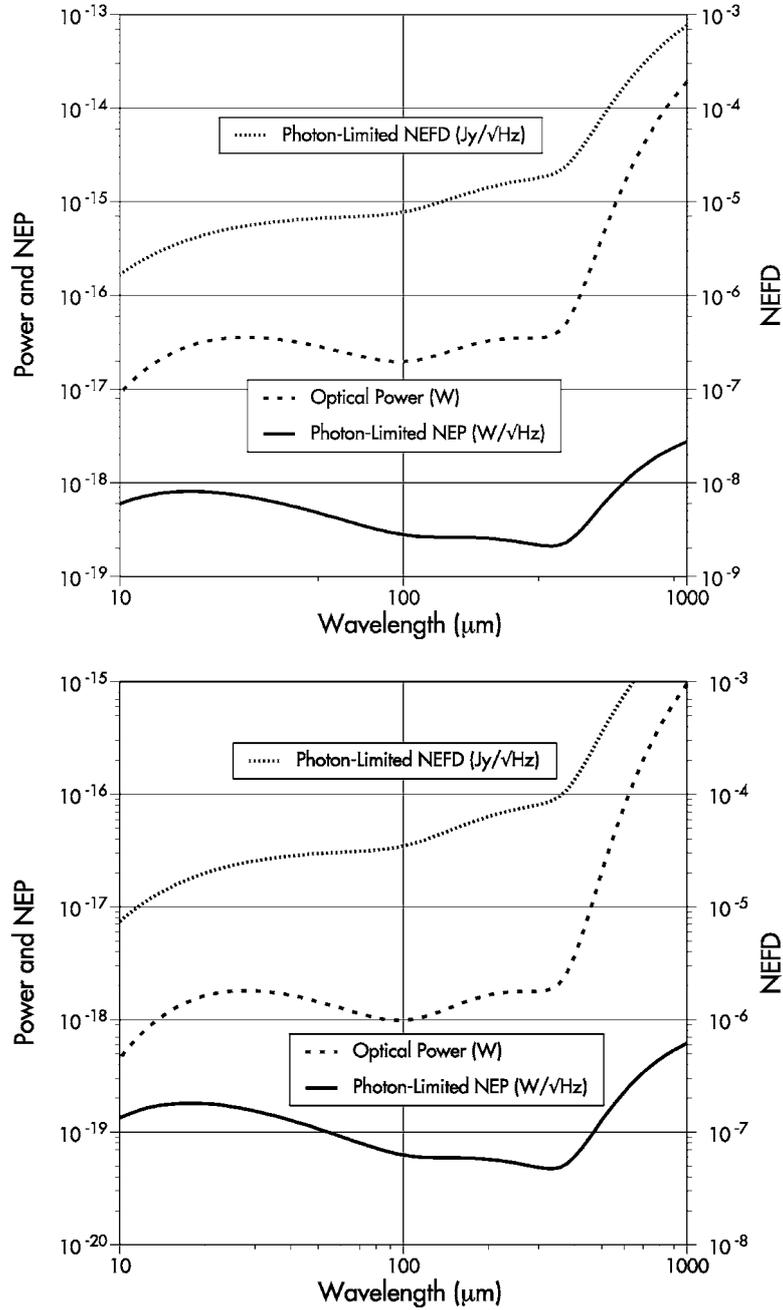

*Figure 16.* (Top) Broadband detector sensitivity: optical power, photon noise equivalent power, and photon-limited noise equivalent flux density for a camera with a spectral resolution of $\lambda/\Delta\lambda = 5$. (Bottom) Optical power, photon noise equivalent power, and photon-limited noise equivalent flux density for a low-resolution spectrometer with a spectral resolution of $\lambda/\Delta\lambda = 100$.

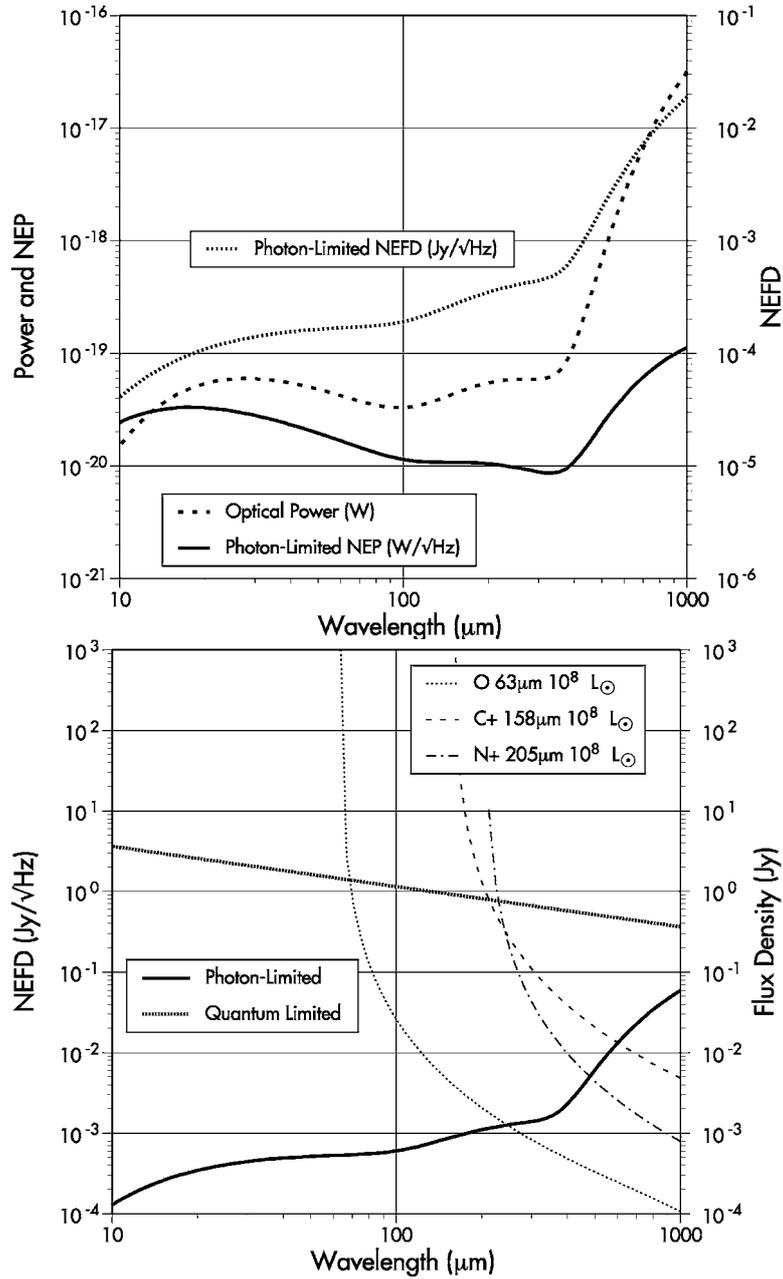

*Figure 17.* (Top) Detector sensitivity: optical power, photon noise equivalent power, and photon-limited noise equivalent flux density for a spectrometer with a spectral resolution of $\lambda/\Delta\lambda = 2000$ (150 km/s). (Bottom) NEFD for a $\lambda/\Delta\lambda = 10^4$ spectrometer using photon-limited direct detectors compared with quantum-limited coherent detectors. Predicted fluxes for redshifted fine-structure lines from galaxies are shown for reference.

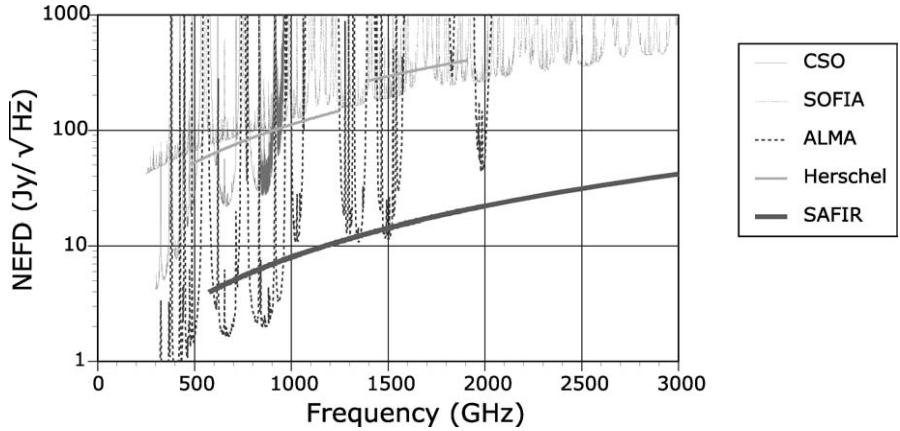

*Figure 18.* Flux density sensitivity at 1 km/s resolution from 300 GHz to 3 THz (100 $\mu$m–1 mm) of several facilities envisioned for the coming decade, as compared to the sensitivity of the SAFIR observatory.

6.4.3. *The Case for High Resolution Spectroscopy*
We compare the predicted sensitivity of a heterodyne spectrometer on SAFIR to similar instruments of other facilities available in the next decade. In Figure 18, we have made an estimate of the NEFD for an advanced instrument suite appropriate to a ground-based 10 m-class telescope (e.g., Kooi et al., 2002), a hypothetical ALMA suite with receivers at all frequencies available from the ground, a hypothetical optimized SOFIA heterodyne spectrometer covering the whole 300 GHz − 3 THz band, and the HIFI instrument on Herschel. NEFD is chosen rather than noise temperature, as the point source sensitivity of each facility is the more relevant parameter. As can be seen, SAFIR will provide a substantial gain in sensitivity at frequencies above 1 THz, enabling new science at these frequencies. Where ALMA is operational, its sensitivity and high angular resolution are superior, but the availability of the >1 THz windows is not common. It is worth noting that all of these facilities are quantum-limited except where the atmospheric emission is strong. For this reason, a warmer but larger implementation of SAFIR would yield better sensitivity for its heterodyne instrument. As mentioned previously, the mirror technology shown in Figure 6 yields a very low areal density but more difficult cooling problems. Thus, it might be more easily scaled to a larger (10–30 m), warmer (∼30 K) telescope than the baseline 10 m, 4 K SAFIR.

### 6.5. THERMAL DESIGN

JWST will use an all-passive design to achieve a telescope temperature of ∼35 K. This is a reasonable practical limit for a telescope relying on radiative cooling alone. Reaching the more challenging 4 K telescope and instrument temperatures requires better isolation from solar radiation and active cooling to get below the ∼7 K ambient (nonsolar) background at L2.

The dominant heat load on the SAFIR observatory is from the Sun; its light must be attenuated by ∼6 orders of magnitude to keep the telescope cold. JWST has designed a sunshade to attenuate this light by ∼3.5 orders of magnitude, using multiple separated radiatively cooled reflective blankets. This sunshade is deployed from the warm spacecraft, and radiatively cools until the inner layer is at a temperature of ∼100 K in its warmest place. This "hot spot" is the dominant source of stray light at mid-infrared wavelengths. For SAFIR, the equivalent "hot spot" must be 15 K or colder, which puts a much greater burden on the sunshade. To meet this requirement, we have mounted a JWST-like sunshade on a 40 K actively cooled stage. The sunshade's sunward side heats up significantly from 40 K, but the inner layer is quite cold. An additional layer is added to the sunshade, mounted on a 15 K actively cooled stage. This layer reaches an equilibration temperature of around 15 K across its entire surface, and thereby prevents stray light from entering the telescope and reduces the radiative load on the cold portions of the observatory to an acceptable level. The salient elements of the thermal design are shown in Figure 19.

Trade studies and thermal analysis have identified a few design features which improve the performance of a JWST-like sunshade. The spacing between sunshade layers will need to be increased slightly as compared to JWST's, to improve radiative cooling of the warmer layers. The extra layer on the cold side is mounted to the telescope tower further up from the spacecraft. The deployment of the sunshade layers will draw on the design used by JWST. The extra inner layer could use a simple separate deployment if necessary.

As previously mentioned, the sunshade is conductively cooled by a 40 K and 15 K mounting point, which is actively cooled by closed-cycle cryocoolers. An additional refrigerator cools the entire telescope and instrument volume (which has its own radiation shield) to 4 K. With the sunshade and cryocoolers in operation, a set of Continuous Adiabatic Demagnetization Refrigerators (CADRs; (Shirron et al., 2002)) is sufficient to cool the instrument and telescope to 4 K and the detectors to the even lower temperatures they require.

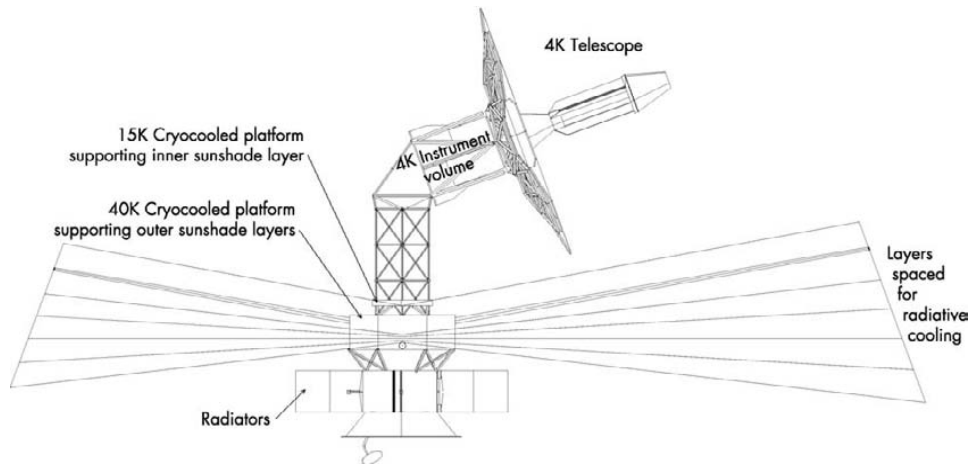

*Figure 19.* SAFIR thermal design, showing temperatures of stages between the warm spacecraft and the 4 K telescope.

### 6.6. Cooling technologies

The current state-of-the-art in active cooling technologies for space use is quite advanced as compared to previously flown, stored-cryogen approaches. This is illustrated in Figure 20, which shows the cooling power available in flight-qualifiable coolers as a function of temperature (adapted from Dipirro, 2003). The requirements for SAFIR are shown as the stars on the plot. None of these cooling requirements is difficult to achieve with modest improvements on existing technologies.

There are several options for space-qualifiable cryocoolers for this temperature range that must be evaluated. The Advanced Cryocooler Technology Development Program (ACTDP) is developing mechanical coolers to provide lifts of 250 mW at 18 K and 7.5 mW at 6 K. For our mission concept, we assumed the ACTDP would develop cryocoolers of the reverse turbo-Brayton cryocooler (RTBC) type, such as the cooler already flying on HST NICMOS. Initial trade studies have led us to baseline the following possible configuration, using the RTBC expected efficiency as a reference point:

– 6 ACTDP-type coolers operating between 300 K and 40 K
– 4 ACTDP-type operating down to 15 K
– 3 Advanced CADRs operating down to 0.05 K and providing telescope cooling to 4 K

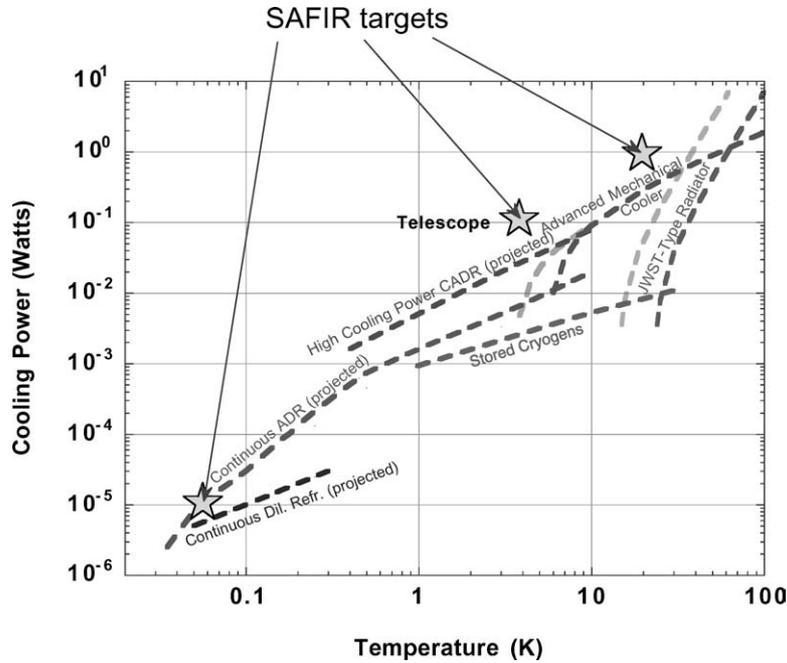

*Figure 20.* Present-day capabilities of cryocooler technologies for space flight purposes.

Very conservative assumptions have been made about the abilities of the Advanced CADRs and the ACTDP coolers. The capabilities needed for SAFIR will likely be demonstrated in the next 2 to 3 years. Although the ACTDP is no longer supporting an RTBC, but rather Stirling-based coolers, the program will develop two coolers with the required capabilities in a near flight-ready configuration by 2005, 10 years before the SAFIR launch. Single shot ADRs have been developed for flight for XRS-1 and sounding rockets and will be flown in 2005 on XRS-2. Continuous ADRs have been demonstrated in the lab operating from warm end temperatures of ∼6 K, and an advanced continuous ADR operating from a 15 K heat sink is believed to be within current manufacturing capabilities. By the time SAFIR enters phase B, both of these coolers should have been demonstrated in flight and be more capable than is assumed in this concept.

The cryocoolers will produce substantial amounts of heat which must rejected at 300 K by radiators mounted on the sunward side of the sunshield. They will be deployed just as the solar panels are, but oriented perpendicular to the sun line.

### 6.7. SPACECRAFT

A large but basic spacecraft bus can accommodate SAFIR requirements. The most challenging task the spacecraft will have to cope with is the control of its relatively massive instrument payload. Maintaining the proper attitude of the observatory—with respect to the Sun and the celestial source under observation—will be a difficult problem, but one which JWST will solve before SAFIR. A combination of instrument-located star tracking and large reaction wheels can solve this problem, as described in Section 6.8. The spacecraft will also house the warm electronics for the instruments and the warm end of the cryocoolers. The power requirements of the coolers pose the second most challenging task. The power handling system for SAFIR is proposed to use a MAP-style system upgraded to 4 kW at 120 VDC operation, supplied by deployed solar panels with a total area of around 17 m$^2$. Cooling is provided by deployed radiators. A large payload attachment fixture is required to handle the high center of gravity and large mass of the SAFIR observatory. The design of the spacecraft bus is shown in Figure 21.

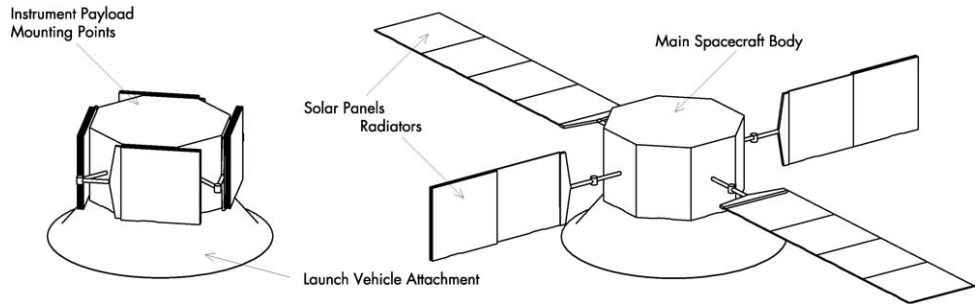

*Figure 21.* SAFIR spacecraft bus (Left) illustrating the large structure required to support the observatory during launch, and (Right) illustrating the deployed solar arrays and radiators.

6.8. POINTING CONTROL

The estimated pointing requirement for SAFIR is to point absolutely to within about 1$''$. This assumes that the smallest field of view is about 5$''$, and the desired point source ought to lie within this region. For a single-beam instrument (HRS), it may not be possible to blind point on faint sources. To remove the expected jitter and drift of the telescope, it is desirable to have relative pointing knowledge to within 0.1$''$, about 1/10th of the smallest diffraction-limited beam. A top-level summary of the possible designs for achieving this is shown in Figure 22. This shows that only two cases (#3 & #5) are suitable. Because #5 requires precision metrology between the warm spacecraft bus and the cold telescope, which are separated by the sunshield, this might be very difficult to implement. Therefore, we believe #3 to be the most credible approach, but that achieving this requires a cold star sensor rigidly mounted to the telescope, so as to provide subarcsecond boresite knowledge.

The conceptual design includes a 2 $\mu$m camera in an unoccupied portion of the focal plane to sense the nearest 2MASS point source (as determined by a spacecraft-mounted warm star tracker) and detect its motion to 0.1$''$ precision to enable shift-and-add image reconstruction for the science data. The 2MASS sources are common and sufficiently bright to make this possible, even if only a fraction of order 10% of the SAFIR aperture is used for this task. Eight momentum wheels are required to meet the requirements on the slew rate of the observatory, assuming redundancy.

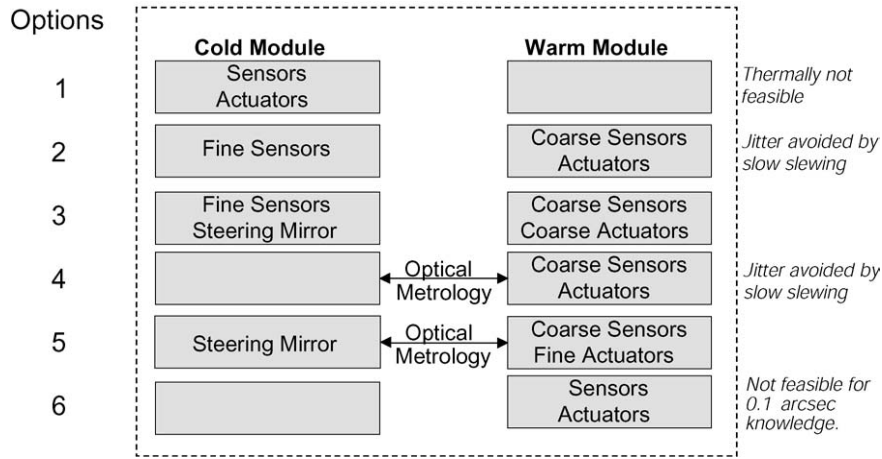

*Figure 22.* Possible designs for the location of sensors and actuators which make up the pointing and attitude control system.

6.9. ORBIT

After studying a variety of orbits, an L2 orbit was found to meet all of SAFIR's requirements for many of the same reasons it meets JWST requirements. This orbit is thermally stable, is always well-illuminated, has a nearly constant distance to the Earth, and the field of regard to space is adequate. The orbit selection is driven primarily by thermal considerations. An L2 orbit allows a large—but feasible— deployed sunshade to create the thermal conditions necessary to allow the telescope to operate at 4 K. The orbit also allows reasonable launch vehicle, communications and power solutions. The path of SAFIR from the Earth to L2 is shown in Figure 23.

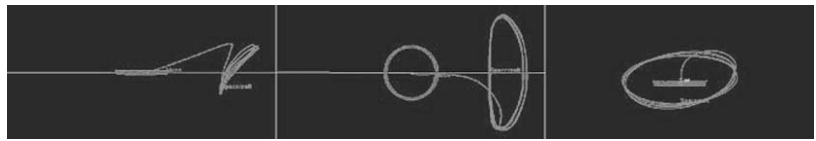

*Figure 23.* Three orthogonal views of the trajectory taken by SAFIR to L2. The orbit of the Moon is the circle near the center.

6.10. COMMUNICATIONS AND OPERATIONS

Calculated data rate estimates are around 800 kbps at most, assuming two instruments observing simultaneously at maximum rate. With an L2 orbit—$1.77 \times 10^6$ km distant—a $\sim$1 m dish using 6 W at X-band can download this data volume to a 34 m DSN base station with one 1.1 h pass per day, with a factor of two reserve on board memory (140 Gb). During the beginning of the mission, sufficient power to run two instruments simultaneously will be available, although toward the end of the mission, degradation in the solar array efficiency will prohibit more than one instrument from operating at a time. Mission operations would be conducted similarly to the JWST operations.

6.11. LAUNCH VEHICLE AND FAIRING

The launch mass of SAFIR is estimated very coarsely at 5000 kg, but this is a conservative mass with substantial margin for growth allowed. The limiting factor for launch vehicle selection, however, is the fairing volume. When stowed, SAFIR will be about 5 m in diameter by 17 m tall (Figure 24; two authors are shown for scale). Shown is the petal-fold architecture, but the use of table-fold or sparse aperture result in a similar volume envelope. All of the deployment and packaging options of our basic concept require a volume only possible with one available fairing, the 19 m fairing on the Delta-IV Heavy. The mass-to-L2 of 9400 kg far exceeds the projected mass of SAFIR. With future refinements, the mass estimate of the SAFIR concept might be reduced enough to fly on launch vehicles with less mass capability, but the fairing size will still not be feasible.

When at L2, the instability of the orbit mandates that SAFIR carry a monopropellant or cold gas system for station keeping. This system must be carefully oriented, such that the propellant not deposit on the cold optics.

6.12. DEPLOYMENTS

JWST will demonstrate most of the deployments needed for SAFIR. The largest deployment will be the sunshade, which for SAFIR will be much like the current JWST designs with one addition. The additional inner layer mentioned earlier will be attached to a stage slightly higher on the telescope tower, cooled to 15 K by active refrigeration. There have been two sunshade deployment concepts investigated for JWST. One concept uses four extending booms to unfurl a thin multilayered blanket. A second concept, which is the selected method, uses sets of unfolding beams with tip spars at the ends to space the various layers. We show both approaches in Figure 4; either is compatible with our telescope designs. The deployment of the telescope—which includes the primary mirror, secondary mirror and telescope cant hinge—will follow the approach of JWST very closely, but perhaps with a simpler phase-up routine allowed by SAFIR's coarser figure requirement. Figure 25 shows one of our SAFIR concepts using the petal deployment approach. The telescope tower for this SAFIR conceptual design is very similar to the JWST telescope tower and deploys in the same way, using the same mechanism carrying with it the cold end of the 40 K and 15 K refrigerators. SAFIR will also deploy a small antenna for transmission of data from the bottom of the spacecraft. On the side of the spacecraft, there are solar panels for power collection and radiator panels for heat rejection, both of which use a conventional panel deployment mechanism.

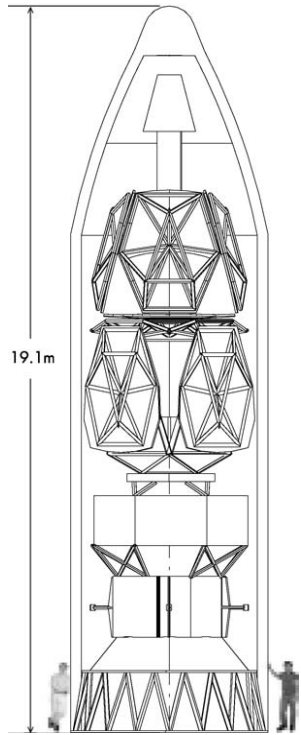

*Figure 24*. SAFIR concept stowed for launch.

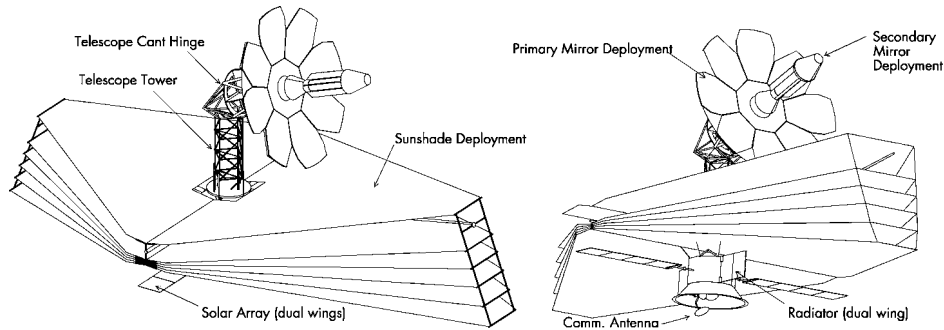

*Figure 25*. Deployments expected for SAFIR.

6.13. INTEGRATION AND TESTING

A great deal of work has already been done to design and plan for JWST integration and testing. Because our primary conceptual design is based largely on JWST, it is not surprising that SAFIR's integration and testing has many of the same challenges. We have made a detailed integration and test plan for the SAFIR concept discussed above, which is quite similar to JWST's plan. The assembly of SAFIR can follow the same basic flow and use many of the same solutions as JWST. Some of the assembly challenges are made easier because of SAFIR's less stringent optical requirements. Some will be made more difficult because SAFIR is larger than JWST.

JWST's testing procedure can also be used as a basis for testing the SAFIR observatory. The purpose of any verification program is to ensure that the required performance of the system will be achieved in flight. The size (10 m vs. 6.5 m) and operational temperature requirements (4 K vs. 35 K) of SAFIR are different from JWST, so the facilities designed for JWST may have to be modified to accommodate SAFIR. JWST will require thermal test capabilities ranging from ambient temperature to liquid helium temperatures. The Plum Brook Facility at NASA Lewis and the Raytheon 50′ diameter spherical thermal vacuum chambers have some promise of being able to accommodate SAFIR testing, with the modifications made for JWST. The basic JWST approach, which uses other facilities designed for the testing of the sunshade and instrument structure, is likely to be viable for SAFIR, with many of the same facilities and setups. "A Strawman Verification, Integration and Test Program For NGST" (1991) describes a JWST integration and test plan including facilities, setups, and justification. Due to the lightweight nature of the JWST telescope and its resultant flexibility, operation in gravity will require the use of carefully designed test supports and self-weight compensators, which will equally apply to SAFIR.

6.14. MASS AND POWER

Running the cryocoolers, instruments, and the spacecraft subsystems requires substantial power. Trade studies have led us to decisions which require more power, but fewer deployments and less design complexity. The power is not hard to obtain given the size of the SAFIR observatory. Solar arrays with total area of $17\,\text{m}^2$ on two standard deployed booms meet the current calculated design estimates with 40% reserve. The estimated power consumption of the entire SAFIR observatory is 3500 W, dominated by around 2000 W for the cryocoolers. This is a very conservative estimate of the total power required, and may be high by as much as a factor of 3. Mass is more difficult to estimate, as very little of the detail needed to do this has been fleshed out. However, we are confident that no major subsystems have been omitted, and so we believe that a conservative estimate of the mass is around 4000 kg. The mass and power budgets are detailed in Tables III and IV.

TABLE IV

SAFIR observatory conceptual design power estimates

| ITEM | Average Power and contingency (W) |
|---|---|
| **PAYLOAD** | |
| CAM (assuming 2 inst on at a time max) | 102 |
|   Contingency | 76 |
| LRS (assuming 2 inst on at a time max) | 20 |
|   Contingency | 15 |
| HRS (assuming 2 inst on at a time max) | 51 |
|   Contingency | 38 |
| HET (assuming 2 inst on at a time max) | 60 |
|   Contingency | 45 |
| Telescope | – |
| Thermal subsystem: | |
|   Cryocoolers (8) | 2000 |
|   Cryocooler radiators (2) | 10 |
|   ADR's (2) | 20 |
|   Sunshield & mechanisms | – |
|   Misc. (blankets, heaters, etc.) | 60 |
| **PAYLOAD TOTAL** | **2499** |
| **SPACECRAFT & OTHER SUBSYSTEMS** | - |
| PSE | 212 |
|   Contingency | 42 |
| Harness Loss | 29 |
|   Contingency | 6 |
| Command & Data Handling | 16 |
|   Contingency | 3 |
| Attitude Control | 220 |
|   Contingency | 44 |
| Com, X-band HGA, HPA, Transponders | 51 |
|   Contingency | 10 |
| Propulsion | 2 |
|   Contingency | 1 |
| **BUS TOTAL** | **1053** |
| **TOTALS (BUS + PAYLOAD)** | **3552** |

6.15. BUDGET

The major systems in SAFIR—for example, the telescope or sunshade—are expected to draw heavily on JWST development and experience. As a result, there will be substantial cost savings for a JWST-like SAFIR. In 1999, Goddard Space Flight Center estimated the budget for SAFIR for the UVOIR panel of the Decadal Survey (NAS Decade Report, 2001) which concluded that the costs would be $310M for construction, $85M for launch, and $100M for 5 years of operations. They drew on their experience estimating the cost of JWST, so the relative comparison of the two missions is also meaningful. They assumed that no additional development would be required beyond that for JWST, although the report indicated that this was probably not entirely correct. A significant development program specific to SAFIR, perhaps even departing significantly from the JWST telescope architecture was assumed to be half that for JWST, or an additional $125M, for a total SAFIR mission cost of $620M. This estimate was made when the best estimate of the UVOIR panel

for JWST was $1114M. The Decadal Survey committee also recommended a budget over the decade for technology development that would support SAFIR (and other projects in the far-infrared and submillimeter) of $140M, including $10M for detectors, $50M for space refrigerators, and $80M for large, lightweight optics.

Our estimate of the SAFIR cost was made by direct line-item comparison to JWST. The result was that SAFIR would cost 80% of JWST. However, this estimate depends strongly on factors that are beyond the program's control, such as its funding profile and prior technology investments.

## 7. Project Status

Investment in the development of the various design concepts is a critical need in the near term. The basic decision of a SAFIR architecture must be made as soon as possible to guide further development of the mission. We have based our concept on JWST, so that its developments may result in approaches that can be readily adapted to the far-infrared, with the differing requirements of (1) colder operating temperature; (2) relaxed image quality; and (3) larger aperture. These three important differences, however, may drive SAFIR to a very different architecture. For instance, although autonomous deployment is a possibility, opportunities for reduced cost and risk through in-space assembly can also be explored.

A team consisting of members from NASA/GSFC, NASA/JPL, and universities has proposed to conduct a deeper study of SAFIR, leading to a more refined mission concept than the one presented here. This will be the dominant SAFIR activity during 2004. Also, there is a high level of interest in the far infrared and submillimeter both in Europe and Japan, making it timely to consider possible international collaborations. Such cooperation has already been fruitful in the infrared and sub-mm for Herschel, Planck, and ALMA.

## 8. Summary

We have developed a conceptual design for the SAFIR observatory, based on JWST's current designs, and made a detailed analysis of the changes necessary to produce SAFIR. The differences between SAFIR and JWST are simple to summarize. Primarily, SAFIR must be much colder, and therefore will require active coolers and a more capable sunshade. The sunshade itself, the largest component of JWST, can be remarkably similar in architecture. The coolers are all feasible using near-future technology. Secondly, SAFIR requires a different complement of instruments, which will take new detectors needing substantial development. Finally, SAFIR's telescope will be larger than JWST's, but the launch vehicle selected can accommodate this change. Other aspects of SAFIR's telescope—the mirror itself, the pointing tolerances—are made simpler by the longer wavelengths and therefore less stringent requirements of SAFIR versus JWST. Table V summarizes the salient parameters of the SAFIR mission.

TABLE V
Mission parameters for achieving SAFIR science goals

| | |
|---|---|
| Primary mirror diameter | 10 meters |
| Wavelength coverage | 20–800 microns |
| Angular Resolution | $1''$ at $\lambda \leq 40\,\mu$m, diff. limited at $\lambda > 40\,\mu$m |
| Science instrumentation | Camera and Spectrographs |
| Mission duration | 5–10 years |
| Orbit | L2 libration point halo, 870 000 km Y amplitude |
| Payload mass | 4900 kg (conservatively, with contingency) |
| Average power | 3600 W (with contingency) |
| Planned launch year | 2015 (approximate) |
| Launch vehicle | Delta-IV Heavy, 19 m fairing |

Funding, technology developments, and the schedule of JWST will all drive the launch date of SAFIR. Based on JWST's 15 year start-to-launch timescale, and assuming some improvement due to JWST's technology developments, SAFIR could launch as early as 2015.

"SAFIR will study the relatively unexplored region of the spectrum between 30 and 300 $\mu$m. It will investigate the earliest stage of star formation and galaxy formation by revealing regions too shrouded by dust to be studied by JWST, and too warm to be studied effectively with ALMA. It will be more than 100 times as sensitive as Spitzer or the European [Herschel] mission. To take the next step in exploring this important part of the spectrum, the committee recommends SAFIR. The combination of its size, low temperature, and detector capability makes its astronomical capability about 100 000 times that of other missions and gives it tremendous potential to uncover new phenomena in the universe." (NAS Decade Report, 2001, pp. 39, 100).

SAFIR will make profoundly important contributions to the goals of both the structure and evolution of the Universe and the origins themes of NASA space science, with clear science priorities and exciting science goals that are intellectually accessible to the greater public. It will also showcase new technology in dramatic advancement for large or cryogenic telescopes, but achieved through realizable technology developments of a moderate scale. We have developed a SAFIR concept at NASA's Goddard Space Flight Center, demonstrating that a JWST-based is a feasible approach to achieving this important mission.

## Acknowledgements

We would like to thank the SAFIR Mission Concept Study Team at NASA/GSFC, for their work in bringing the concept detailed above to completion. They include: Ed Canavan, Dave Green, Cathy Marx, Mark Matsumura, Lee Mundy, Keith

Parrish, Joe Pelliciotti, Juan Roman, and Kurt Ruckleshaus. Guidance for the science requirements of this concept, and some scientific background information, resulted from substantial contributions from George Rieke, Paul Harvey, Dan Lester, and Lee Mundy.